\documentclass[prc,twocolumn,showpacs,preprintnumbers,amsmath,amssymb]{revtex4}
\usepackage{graphicx}% Include figure files
\usepackage{dcolumn}% Align table columns on decimal point
\usepackage{bm}% bold math
\usepackage{amssymb, amsmath}

\newcommand{\iso}[2]{$^{#1}\text{#2}$}
\newcommand{\bor}[0]{\iso{8}{B} }
\newcommand{\nit}[0]{\iso{12}{N} }
\newcommand{\gs}[0]{$\rm {^{8}B(2^+)} \rightarrow {^{8}Be(0^+)}$ }
\newcommand{\es}[0]{$\rm {^{8}B(2^+)} \rightarrow {^{8}Be(2^+)}$ }

\begin{document}

%\preprint{APS/123-QED}

\title{ Search for the second forbidden beta decay of $\bm{^8}$B to the ground state of $\bm{^8}$Be }% Force line breaks with \\

\author{M. K. Bacrania}
 \altaffiliation[Presently at ]{Nuclear Safeguards Science and
   Technology, Los Alamos National Laboratory, Los Alamos, NM 87544}%Lines
                                %break automatically or can be forced
                                %with \\
\author{N. M. Boyd}
\altaffiliation[Presently at ]{Department of Physics and Astronomy,
  University of British Columbia. Vancouver, BC V6T 1Z1}
\author{R. G. H. Robertson}
\author{D. W. Storm}%
\affiliation{%
Center for Experimental Nuclear Physics and Astrophysics,
Physics Department,
University of Washington,
Seattle, WA 98195
}%

\date{\today}% It is always \today, today, 
% but any date may be explicitly specified

\begin{abstract}

A significant decay branch of $^8$B to the ground state of $^8$Be would 
extend the solar neutrino spectrum to higher energies than anticipated 
in the standard solar models.  These high-energy neutrinos 
would affect  current neutrino oscillation results and  also would be a background to 
measurements of the hep process. 
We have measured the  
delayed alpha particles from the decay of $^8$B, with the goal of observing 
the two 46-keV alpha particles arising from the ground-state decay. 
The $^8$B was produced using an in-flight radioactive beam technique. It was 
implanted in a silicon PIN-diode detector that was capable of identifying the
alpha-particles from the $^8$Be ground state.  From this measurement we find
an upper limit (at 90\% confidence level) of $7.3 \times 10^{-5}$  for the branching 
ratio to the ground state.  In addition to describing this measurement, we present a theoretical
calculation for this branching ratio.

\end{abstract}

\maketitle

\section{\label{sec:intro}Introduction}

There have been several experiments \cite{freedman, ega, ortiz} done recently to measure
the alpha-particle spectrum following the allowed ($2^+$ to $2^+$)
beta decay of $^8$B to the broad first excited state at 3 MeV in
$^8$Be.  The neutrinos from this decay provide the signal for solar
neutrino measurements, such as SNO and SuperKamiokande, and precise
knowledge of their spectrum at creation, combined with accurate
measurements in the solar neutrino detectors, is important for
understanding neutrino oscillations.  The only higher energy neutrinos
from the Sun come from the hep \cite{SNO-hep} process (weak capture of a proton by
%added citation per Laura #2
$^3$He producing $e^+ + \nu$), expected to be at a rate roughly
$10^{-3}$ of the $^8$B neutrinos, and from the second forbidden ($2^+$
to $0^+$) decay of $^8$B to the ground state of $^8$Be.  Specifically these reactions are
\begin{equation*}
^3\text{He} + p \rightarrow ^4\!\text{He} + e ^+  +  \nu  + 18.77  \text{  MeV }
\end{equation*}
and 
\begin{equation*}
 ^8\text{B}  \rightarrow   ^8\!\text{Be} + e^+ + \nu + 16.95 \text{ MeV. }
\end{equation*}

 An $ft$ value for the latter decay, estimated using $\log_{10}ft = 13.3$
obtained from  
the $2^+$ to $0^+$ decay of $^{36}$Cl, suggests the $^8$B $2^+$ to
$0^+$ decay should have a branching ratio of order only $
10^{-7}$. The $^8$B decay has an unusually large energy, and
second-forbidden decays should have an extra factor of $k^2$ ($k$ is
the momentum transfer to the leptons) in their decay rate.  Comparing
the square of the endpoint energy for $^8$B with that for $^{36}$Cl
suggests an additional factor of 570 in the rate for the second
forbidden decay of $^8$B.  This rate would correspond to a
$\log_{10}ft$ of 10.5, a value still in the expected range for
second-forbidden transitions.  The corresponding branching ratio would
be $3 \times 10^{-5}$.  If the branching ratio were as large as about
1\%, it would contribute significantly to the neutrino spectrum
observed at SNO. 

%break for new paragraph
 Decay to the ground state of $^8$Be is difficult to
measure, since that state decays to two alpha particles producing only
92 keV.  We are unaware of previous attempts to measure this
 branch.
Published estimates range
 from $3 \times 10^{-6}$ \cite{tribbleest}
to $O(10^{-4})$ \cite{ortizthesis, garciapriv}.  The latter estimate
includes only the vector contribution to the second-forbidden decay.
In this paper we describe the technique of our measurement, give the
result, and present theoretical calculations based on the shell
model and on measured gamma-ray transition matrix elements.

\section{\label{sec:expt}Experiment}

We made a  \iso{8}{B} beam using an in-flight production technique.  The \iso{8}{B} were
implanted in a silicon detector that was capable of measuring the energy from the delayed
alpha decay of the \iso{8}{Be} ground state, as well as from most of the spectrum of
delayed alpha decay of the \iso{8}{Be} first excited state.  The decays were tagged by 
detecting the decay positrons in a scintillator.  The silicon detector was calibrated using 
back scattered $\alpha$ particles of known energy and also using delayed alpha decays of
implanted \iso{12}{N}.

\subsection{\label{sec:radbeam}Radioactive beams}

The $^8$B for implantation in the detector was made using the
$^3$He($^6$Li,$^8$B)n reaction.  A 32-mm-diameter gas cell was located
at the focal point of the analyzing magnet of the University of
Washington Tandem accelerator.  This gas cell had entrance and exit
windows of 2.5-$\mu$m Havar \cite{havar} and contained $^3$He at
about 0.6 bar.  The incident beam was 24-MeV $^6$Li, which, after
accounting for energy loss in the windows and gas, produced $^8$B in
the forward direction with an energy centered at 15.5 MeV and a spread
of 1.1 MeV, resulting from energy loss in the gas.

The beam transport following the analyzing magnet consisted of a
$30^\circ$ switching magnet followed by a quadrupole doublet.  These
elements are normally used to focus the analyzed beam from the
accelerator onto the target in the scattering chamber.  For the
present experiment we had to optimize the capture and transport of
the radioactive beam through the 20-mm aperture for our detector,
which was located at the target position in the scattering chamber.
 We found, using {\scshape transport} \cite{trans} and {\scshape turtle}
\cite{turtle}, that this could be done by first tuning
the beamline following the gas cell 
appropriately for transport of a straight-going monoenergetic beam of
the desired magnetic rigidity, and then by increasing the quadrupole
strength by about 10\%.
The energy acceptance of this system was only about 2\%, which is less
than the energy spread resulting from the gas cell, so there was no
reason  
to have either a bigger cell or denser gas.  We tuned the
radioactive beam transport elements by tuning first a 7.44-MeV
$^6$Li$^{3+}$ beam (matching the 15.5-MeV $^8$B$^{5+}$ radioactive
beam in magnetic rigidity) through a small aperture. Then, while %Laura change
irradiating the gas cell with the 24-MeV $^6$Li beam, we
would monitor the $^8$B implantation rate and increase the quadrupole
current to maximize the rate.  We also swept the switching magnet
field, which we monitored with a Hall probe, to make sure we were
accepting the central energy from the gas cell.

As the rigidity of the primary beam is higher than that of the
radioactive beam, degraded primary beam is a serious contaminant.  In
order to minimize degradation of the primary beam after the analyzing
magnet, we regulated the accelerator voltage with the Generating Volt
Meter Regulator \cite{gvm} and completely retracted the beam
regulation slits located downstream of the analyzing magnet.
Particular attention had to be paid to the beam tuning, in order to
minimize the amount of beam 
striking other limiting
apertures in the beamline. 

In addition to the $^8$B beam, we produced a $^{12}$N beam for
calibration of the silicon detector.  This beam was made using the
$^3$He($^{10}$B,$^{12}$N)n reaction, with a 35-MeV $^{10}$B beam,
producing 24.2-MeV $^{12}$N$^{7+}$, in an analogous manner.  The
$^{12}$N energy was chosen to produce the same implantation depth for
the $^{12}$N as for the $^{8}$B. The choice of the $^{12}$N charge
state was determined by the background, 
because the most-probable
charge state of $^{12}$N, $6+$, was too close in rigidity to the
\iso{10}{B} primary beam. Using the coincidence with the scintillator,
we were able to tag about $3-5\ ^8$B per second with 170 to 330 particle
nA of $^6$Li, and about $0.5-1\ ^{12}$N per second with 50 to 100 particle
nA of $^{10}$B.  For the \iso{8}{B} beam, the background in the silicon
detector, 
 primarily
consisting of degraded \iso{6}{Li} from the primary beam, 
was $0.6-1$~kHz. For the \iso{12}{N} beam, the degraded \iso{10}{B} background was
2 kHz.  This background rate in our detector set an intensity limit
for the primary beam.

\subsection{\label{subsec:detector}Detector}

The detector assembly consisted of a 500-$\mu$m thick by 18-mm square
Hamamatsu S-3204-06 PIN diode \cite{pin}, in front of which was a
51-mm cube of Bicron-400 scintillator with a 20-mm-diameter hole on
the beam axis.  A lead shield 6.4-mm thick with a 17-mm-diameter
aperture was placed in front of the scintillator, and the scintillator
was coupled on the side to a photomultiplier tube. 
This thickness of lead was sufficient to absorb positrons from
decaying $^8$B that 
stopped on it.  Ions that passed through the lead
aperture would stop in the PIN diode, and were not able to hit the
sides of the hole in the scintillator.

\begin{figure}
\includegraphics[width=.49\textwidth]{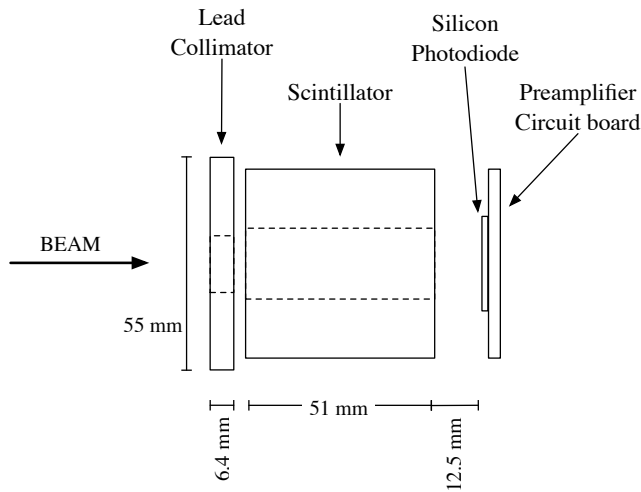}
\caption{\label{fig:detector} Schematic of the detector assembly. The various hole diameters are given in the text. The cooled copper mount is not shown, but was located behind the preamplifier circuit board.  
}
\end{figure}

The 15-MeV $^8$B ions were implanted approximately 20 $\mu$m into the
silicon, and about 20\% of the decays produced positrons that would
enter the scintillator, tagging the decay. 
%    add following bit
In order to minimize distortions in the $\alpha$-particle spectrum from positron energy deposition, the scintillator  detector was placed in front of the silicon detector, thus minimizing  the track length of the tagged positrons in the silicon detector.  Tagging with positrons that passed through the silicon detector was not practical, as their energy deposition would be larger than the energy from the decay from the \iso{8}{B} ground state.  Indeed, as is discussed in section~\ref{subsec:recoil}, positrons that scatter through large angles affect the efficiency of the detection.
The PIN diode was mounted
on a custom charge-sensitive preamplifier \cite{preamp} built on a
5-cm-square circuit board. The preamp and detector were
cooled by a thermoelectric cooler 
coupled to a 19-mm-wide and 3-mm-thick copper strip that ran the
height of the preamplifier circuit board.  This assembly is shown in
Fig.~\ref{fig:detector}.

When cooled to $10^\circ$ C, the measured leakage current of a new PIN
diode was typically $10-12$~nA.  The resolution (FWHM) was 7.8 keV for
81-keV gamma rays, and a threshold about 30 keV was practical.  During
experimental runs, the reverse current would rise to $30-60$~nA after 2
weeks of irradiation, at which point the resolution would decrease to
approximately 10 keV at 81 keV.

\subsection{\label{subsec:elect}Electronics and data acquisition}
The photomultiplier on the scintillation detector produced both a
timing and a pulse-height signal.  The timing signal started a time to
amplitude converter (TAC). The output of the charge-sensitive
preamplifier for the silicon diode detector was split into two
channels, each with an Ortec 572 amplifier.  The maximum range of the
{\em high gain} channel corresponded to about 1 MeV, while the maximum
observable energy in the {\em low gain} channel was 8 MeV.  The two
unipolar signals were digitized, as was the pulse-height signal from
the scintillator.  The bipolar outputs of the amplifiers for the
silicon detector generated timing signals, which were combined in an
OR module, and the resulting logic signal was delayed and used to
trigger the data acquisition system (DAQ), contained in a single computer-automated
measurement and control (CAMAC)
crate.  A parallel output of the OR unit was used to stop the TAC,
which was started by the signal from the scintillator.  This TAC
signal was also digitized.  Thus any signal over the 30-keV threshold
from the silicon detector would trigger the DAQ, while a coincidence
(within a 5-$\mu$s time window) would also produce a TAC output.  The
threshold on the scintillator was set high enough that its rate was
substantially lower than that in the silicon diode detector.  The DAQ
system was operated in event mode, with two silicon detector energy
signals, one scintillator pulse height, and one timing signal per
event.

A precision pulser, coupled to the silicon detector preamplifier, was
used for monitoring dead time and amplifier gain shifts.  A scaler
module was used to count pulser triggers, pulses from the beam-current
integrator, the number of TAC outputs, and the number of DAQ triggers.
The pulser and the electronics feeding the DAQ system were located in
a temperature-stabilized rack, which was controlled to within $\pm
0.1^\circ$C.  During regular data taking the pulser was set to
correspond to about a 0.5 MeV signal in the silicon detector.  The DAQ
CAMAC crate was located in the temperature-controlled electronics
rack and was controlled by a Wiener CC32 peripheral component interconnect 
(PCI) crate controller
\cite{wiener}. For data acquisition, initially we used the {\scshape java}-based
{\scshape jam} \cite{jam} system, but subsequently switched to the
Objective-C-based {\scshape orca} \cite{orca} system.

The gain stability of the silicon-detector preamplifier and associated electronics
was monitored by observing the pulser signal in the high- and low-gain
channels. Also, the irradiations were stopped at hourly intervals, and
a measurement of the pulser spectrum for various settings of the
pulser attenuation circuit was made $(1.0\times, 0.5\times, 0.2\times, 0.1\times)$. These
measurements allowed for the monitoring of both gain and linearity as
a function of irradiation time.

From the event-mode data, singles histograms of the silicon-detector
energy spectra, the TAC signals, and the scintillator pulse height
were formed.  Prompt coincident events were retained   
 by applying cuts (discussed in section~\ref{sec:analysis}) to
the TAC and scintillator spectra.

\subsection{\label{subsec:calib}Calibration}
\subsubsection{Silicon detector energy calibration}

The energy calibration for the \gs decay measurement 
was obtained from a \iso{133}{Ba} radioactive source, which produces a
number of low-energy gamma-ray lines between $80-400$~keV.  These
photons convert by Compton scattering and the photoelectric effect in
the silicon.  In contrast, measurement of the \gs decay rate 
involved 
the detection of two low-energy alpha particles, which directly ionize
the detector medium. 
The ionization energy for gamma rays in silicon,
$3.68\pm0.02$ eV, is larger than the ionization energy for alpha
particles in silicon, $3.62\pm0.02$ eV. This difference, a ratio of
$1.017\pm0.008$, must be accounted for when relating a calibration
obtained from gamma-rays to the measured alpha-particle energy
spectrum \cite{ICRU:1979fw}.  
In addition, the fractional loss of energy to non-ionizing
processes increases as the alpha particle incident energy
decreases; this {\em pulse-height defect} goes undetected by the
silicon detector.  

Tabulated values of the expected nuclear and total
stopping powers have been published in \cite{Berger:wn}. At low
energies, these tabulated nuclear stopping powers are calculated via a
model, 
as available experimental data are limited. In order to
determine the expected 
pulse height of the \gs decay signal reliably, it was
necessary to measure 
the pulse-height defect of alpha
particles in our silicon detector.

The uncertainty in the ionization energy ratio affects the
determination of the pulse-height defect, but, as will be seen below,
the calibration of the low-energy alpha particles is not affected by
it. Thus we divide the calibration obtained from \iso{133}{Ba} by 1.017 to get 
the corresponding calibration for alpha particles without the pulse-height
deficit, and these energies are used for our calibrations.
The pulse-height deficit (with uncertainties) is applied subsequently.

In order to measure the pulse-height defect of low-energy alpha
particles in silicon, \iso{4}{He} ions were produced at eight
different energies from $90-475$~keV using the UW Tandem accelerator
with the Terminal Ion Source. These alpha particles were backscattered
from a gold monolayer evaporated on a 50-$\mu$g/cm$^2$ carbon target foil
into the collimated silicon detector located at 100 degrees. As the
energy the particles deposit in the active part of the detector is
reduced from the incident energy by the SiO$_2$ deadlayer on the front
of the detector, it was necessary to quantify the energy loss in the
deadlayer.  This was done by repeating the measurements for each
incident energy with the detector face angled at numerous positions
between $-45$ and $+45$ degrees to the scattering axis. At each angle, the
effective thickness of the deadlayer through which the particles
passed is $\Delta x/\cos{\theta}$, where $\Delta x$ is the physical
deadlayer thickness.  The value of $\Delta x$ was determined by
combining all of the measured results and performing a global fit to
the measured energy (as determined from the \iso{133}{Ba} calibration,
adjusted by the ratio of ionization energy for electrons and alpha
particles) vs.~detector angular position. From this series of
experiments, the pulse-height defect was found to vary from $7.9 \pm
0.3$ to $11.4 \pm 0.7$ keV for alpha-particle incident energies in the
active part of the detector varying from 55.9 to 391.4 keV.  The
measured energy loss in the deadlayer corresponded to $43.1 \pm 0.4 
\mu$g/cm$^2$ of SiO$_2$.  There is an additional systematic
uncertainty of 0.8\% in these pulse-height defects, arising from the
uncertainty in the ratio of ionization energy for alpha
particles to that for photons \cite{ICRU:1979fw}. 

The low-energy alpha calibration, including corrections for 
pulse-height defect and for the ratio of ionization energy for electrons and
alpha particles, was cross-checked using the decay of implanted
\iso{12}{N}, which beta-decays to the 7.65-MeV state of \iso{12}{C}, 
which then decays to three alpha particles with a total energy
release of 379 keV.

\begin{figure}

\includegraphics[width=.48\textwidth]{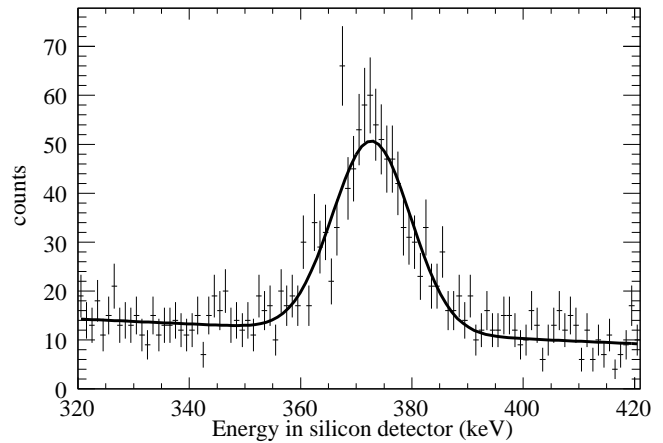}
\caption{\label{fig:12N}The \iso{12}{N} $3\alpha$ decay peak observed
in the high gain channel, fit with a Gaussian peak on a linear background.
The energy deposited, including recoil and positron summing, is predicted 
to be $388.5 \pm 1.2$ keV from our Monte Carlo simulations.  The center of 
the peak appears at $372.5 \pm 3.0$ keV, indicating a total $16 \pm 3$ keV 
pulse-height defect.}
\end{figure}

\begin{figure}
\includegraphics[ width=.48\textwidth]{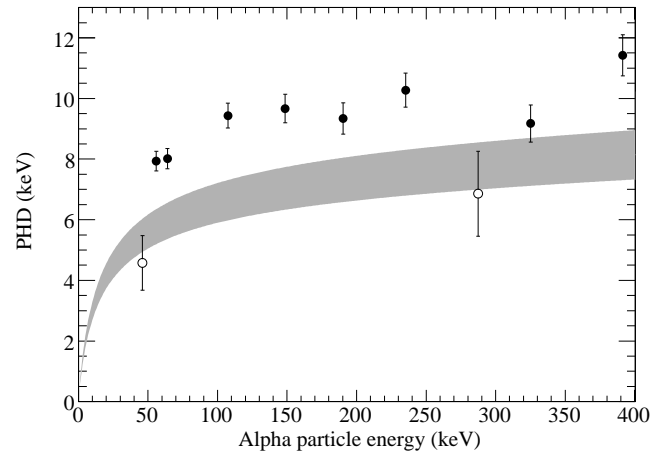}
\caption{\label{fig:phd} Pulse-height defect (PHD) measured using single
  alpha particles (solid circles) and from \iso{12}{N} decay (open
  circles). The shaded band is the PHD value obtained from integrating
the stopping-power tables in \cite{Berger:wn}, including a 10\%
uncertainty. }
\end{figure}

Two of the alpha particles come from the breakup of the ground
state of \iso{8}{Be}. By comparing the known energy of this decay
(accounting for decay kinematics, discussed in the following section)
to the detected energy deposition, we were able to determine the
pulse-height defect independently, assuming that the energy dependence
of the pulse-height defect was that given by the predicted
non-ionizing stopping power \cite{Berger:wn}. The spectrum obtained from the \iso{12}{N} decay is shown in Fig.~\ref{fig:12N}.  The results of the
\iso{12}{N} decay and the single alpha-particle measurements are shown
in Fig.~\ref{fig:phd}, along with the predicted pulse-height defect
obtained from \cite{Berger:wn}. The pulse-height defect obtained from
the single alpha-particle measurement is calculated to be $7.5\pm0.3$
keV at 46 keV. On the other hand, the pulse-height defect from the
\iso{12}{N} decay measurement is found to be $4.5\pm0.9$ keV at 46 keV. 
We use
these two values as lower and upper bounds when constructing the
energy window for the expected signature from \bor decay.

As the low-energy alpha calibration includes corrections for both
pulse-height defect and for the ratio of ionization energies, and as
the former is obtained using the latter, uncertainties in this ratio
will cancel in the calibration for the \iso{8}{B} measurement.  
The photons are just an
intermediate calibration.  The determination of the pulse-height
defect links a known alpha-particle energy to a photon energy, and
then the use of that photon for calibrating the measurement of
\iso{8}{B} establishes that calibration in terms the known
alpha-particle energy.

\begin{figure}

\includegraphics[width=.48\textwidth]{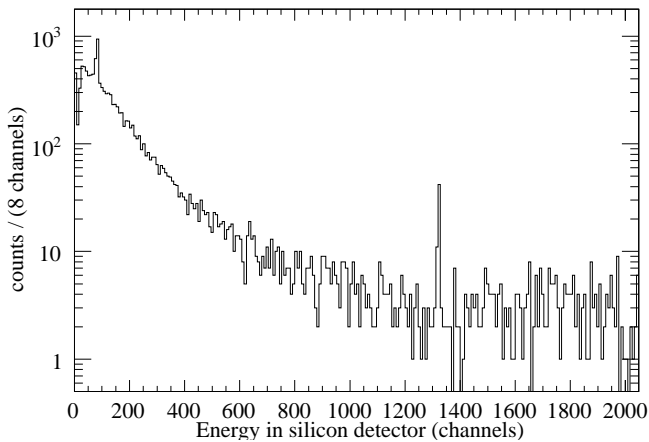}
\caption{\label{fig:12N2peak}The  \iso{12}{N} spectrum observed
in the low-gain channel.  The peaks near channels 80 and 1300
correspond to total  
energies from $3\alpha$ decay of $388.5\pm 1.2$ and $5443 \pm 6$ keV, 
respectively, including recoil and positron summing. The continuum is dominated by positrons backscattering from the detector mount and thus passing through the detector twice. In \iso{12}{N} decay, only 3\% of the positrons are accompanied by alphas. }
\end{figure}

The low-gain channel, which contained the $^8$B$(2+) \rightarrow ^8$Be$(2+)$ 
decays, was calibrated using the two lines from \iso{12}{N} decay, resulting 
from the $3\alpha$ decay of the \iso{12}{C} levels at 7.65 and 12.71 MeV. These
are illustrated in Fig.~\ref{fig:12N2peak}.  The 
energies of these lines were adjusted for positron summing, \iso{12}{C} 
recoil and pulse-height defect.

\subsubsection{ \label{subsec:recoil}Recoil and energy deposition calculations}

Two corrections are required for the analysis of both \bor and \nit
energy spectra: the recoil energy imparted by the leptons to the
daughter nucleus, and the amount of energy deposited in the silicon
detector by the outgoing positron. As a result of the high decay energy, a
significant amount of recoil energy (up to 19 keV) can be imparted to 
the \iso{8}{Be} daughter nuclei by the outgoing leptons, resulting in
an overall broadening of the energy window in which the decay signal
is expected to lie. The   
\iso{12}{C} daughters receive less recoil energy (up to 3.3 keV). 
Using the predictions of the positron-neutrino correlation from the calculations 
described in Sec.~\ref{sec:theory}, we found a distribution of recoil 
energy between 0 and 19 keV for the \gs decay, which was peaked near the maximum energy.  
The Gamow-Teller decay of \nit yields a fairly uniform distribution of recoil 
energy between 0 and 3.3 keV.

The additional energy deposited by the escaping positron adds to the
decay energy and results in an overall energy shift. We minimized the
magnitude of this energy shift by implanting the \bor and \nit ions
near the upstream surface of the silicon detector. The distribution of
the positron energy deposition was computed by Monte-Carlo simulation,
using measured stopping powers from \cite{Berger:wn}. The \bor and
\nit implantation depths, $20.2 \pm 0.3$
and $18.7\pm1.5$ microns respectively, were obtained from 
\cite{Ziegler:lb}. The simulated energy distributions were fit with a
Landau distribution, and the most likely energy deposition for
positrons coming from \gs and \nit decay was $9.5\pm0.9$ and
$7.8\pm0.6$ keV respectively.

A significant number of positrons undergo large angle scattering, in  either
the silicon or  the copper detector mount, before reaching the
scintillator
and resulting in a coincidence trigger. Their energy
deposition in the silicon detector is large enough to remove 
the event entirely from the expected
energy range for the \gs decay.  
As described below, we determined that positron summing reduced our 
efficiency for detecting ground state decays by 11.5\%.

%new section%%%%%%%%%%%%%%%%%%%%%%%%%%%%%%%%%%%%%%

\section{\label{sec:analysis}Data analysis}

Following a series of commissioning runs, we accumulated
$2.5\times10^{6}$ \bor decay events  over the course of two
experimental runs. After selecting prompt events and applying cuts  on the amplitude of the pulse in the scintillation counter, 
$2.0\times10^{6}$ events remained.  The following discussion applies to both runs and 
is illustrated by examples from the second.

Following each run, the data in each of the silicon detector channels
were corrected for gain drifts (the average correction was less than
0.3\%), and the high-gain channel was calibrated using the information
from the \iso{133}{Ba} calibration data. A correction was also made
for the alpha/gamma ionization energy difference. No adjustments or
calibrations were made to the scintillator or TAC spectra.     
The events of the final data set consisted 
of the corrected silicon-detector values and the corresponding raw
scintillator and TAC values. 
This processing was applied separately to
the data obtained using the \iso{3}{He} and \iso{4}{He} target gas.
The ratio of 
random to true coincidence rate, in the energy region where the ground state
decay is expected, was found
to be approximately 0.5\% for both the \iso{3}{He} and \iso{4}{He} data.
Since these data were subtracted to obtain the final data set, we made no
further correction for accidental coincidences.

\begin{figure}

\includegraphics[width=0.49\textwidth]{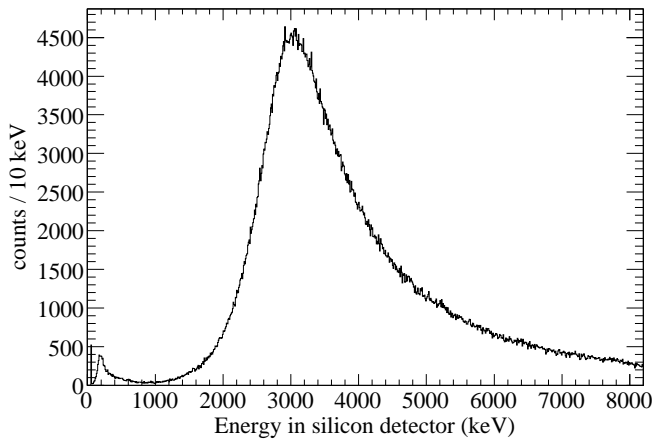}

\caption{\label{fig:lowgain} Spectrum from the low-gain channel of the silicon
detector, dominated by
the $^8$B$(2+) \rightarrow ^8$Be$(2+)$ decay.  As discussed in the text, 
the 20-$\mu$m implantation depth limits the reliability of this spectrum
to energies below 8 MeV.  }

\end{figure}

The spectrum from the low-gain channel, which is dominated by  
the $^8$B$(2+) \rightarrow ^8$Be$(2+)$ decay, is shown in Fig.~\ref{fig:lowgain}.
Above about 8 MeV, Monte Carlo simulations indicate the spectrum is distorted
by alpha particles that escape from the 20-$\mu$m implantation depth.
Additional small distortions result from the  scintillator threshold, which 
requires positrons over about 1.5 MeV,  and
consequently favors lower energy alpha particles.
Finally, significant distortions result from backscattered positrons, many of which 
come from the copper cooling mount behind the detector.  Presumably as a result of these 
distortions, which are examined in detail in Ref.~\cite{mineshthesis}, we were 
unable to get a useful R-matrix fit to this spectrum.

\begin{figure}
  \includegraphics[width=.48\textwidth ]{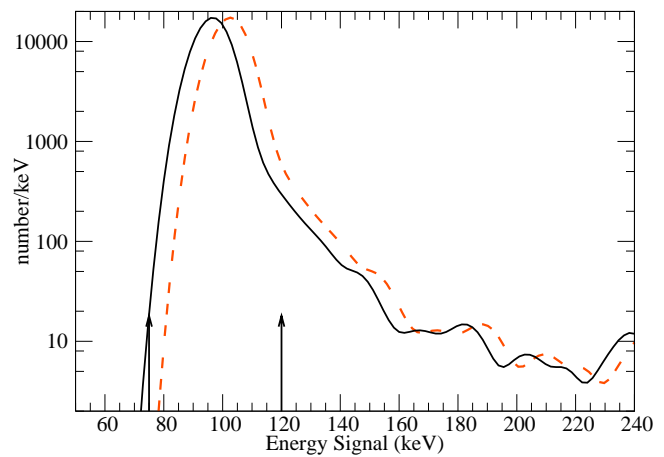}
   \caption{\label{fig:pspect} Monte Carlo simulations of silicon  detector response
	to decays to the \iso{8}{Be} ground state
      for coincident events in scintillator and Si  detectors. The initial 
     alpha-particle energy spectrum is modified by  recoil and positron energy 
     summing. The two curves correspond 
     to the two extreme values for the pulse height defect.  The arrows 
     indicate the region chosen for searching for ground state decays. 
}
\end{figure}

To determine the energy window in which the \gs decay should be
located, we begin by adding the 
recoil energy, distributed 
between 0 and 19.3 keV,  
to the total kinetic energy (91.8
keV) of the two alpha particles arising from \iso{8}{Be} decay. To
calculate the pulse-height defect of these alpha particles, we assume
that the total energy is shared equally between the two alpha
particles, thus giving a minimum and maximum kinetic energy for each
alpha particle of 45.9 keV and 55.6 keV, respectively. The range of
the pulse-height defect for these alpha particles was measured to be
$4.5-7.5$~keV.  Thus, before accounting for the positron energy loss,
the ground state decay could be expected to
appear between 9 and 15 keV lower than the sum of the decay and recoil 
energies.
We convolve the calculated alpha-particle energy distribution with
a Gaussian detector response with the typical 9-keV FWHM to produce an expected energy spectrum.
The pulse height defect shifts this  spectrum downward between 9 and 15 keV.  For each of these cases 
we convolve the shifted spectrum with the spectrum of positron energy deposition 
calculated using the {\scshape Penelope} Monte-Carlo package 
\cite{J.-Sempau:1997xs}.  That spectrum corresponds to positrons emitted isotropically but 
depositing over 2.5 MeV in the scintillator.  The spectrum has a peak at about 9 keV, corresponding to positrons that exited the silicon in the direction of the scintillator, but there is a long tail produced by 
positrons that either backscattered from the copper mount or that passed through 
significant thickness of silicon before a large-angle scattering into the scintillator.
The final spectra, for the two limiting values of pulse height defect, 
including positron summing, are illustrated in Fig.~\ref{fig:pspect}.
We select the energy range from 75 to 120 keV for the region of interest. This region contains between 88.1\%
and 88.8\% 
of the total events, depending on the assumed value for the pulse height defect.

% penelope spectra

In this energy window the rates of the two background processes in \iso{8}{Be},
gamma-decay of excited states of \iso{8}{Be} and alpha particles from
the broad \es decay spectrum, were calculated to be two orders of
magnitude below the sensitivity of this experiment.

\begin{figure}
\includegraphics[width=.49 \textwidth]{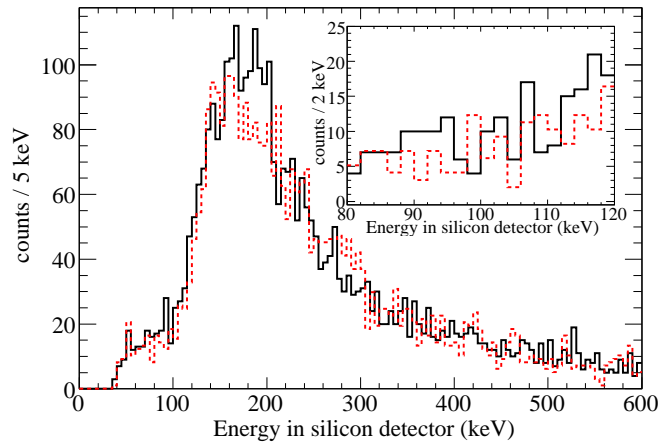}
\caption{\label{fig:le_run5} Measured signal (solid) and background
  (dashed)  coincidence silicon-detector spectra for the high-gain channel with
  scintillator threshold applied. }
\end{figure}

The spectra obtained in the second experimental run, with \iso{3}{He} (signal) 
and \iso{4}{He} (background) in the gas cell,  
are shown in Fig.~\ref{fig:le_run5}. 
The most
striking feature is the large broad peak of coincident events, which
was present in our data regardless of the target gas, and (at a
slightly lower rate) even without the beam. 
 The rate of events in this peak, 20 mHz, was relatively constant over a long time span ($\sim4$
years), regardless of the time between irradiations in the scattering
chamber. We speculate that this peak is due to minimum-ionizing events
from ambient radioactivity which pass through both the silicon and
scintillator detectors. However, increased shielding and the addition
of scintillator veto counters were not effective in reducing the rate.

Analysis of these events shows that they are dominated by low-energy
scintillator events, and we can improve our experimental sensitivity
by optimizing the scintillator minimum threshold.  The optimum
scintillator threshold was obtained by finding the minimum uncertainty in the
difference of the time-normalized signal and background counts between
$75-120$~keV as a function of scintillator threshold.  
We maintained the
average beam current in the measurements with \iso{3}{He} and \iso{4}{He} to be identical
within 1\%, so normalizing by time was equivalent to normalizing by integrated current.
We subtracted the normalized background given by the \iso{4}{He}
data from the \iso{3}{He} data.

Along with the  
background discussed above, there was additional background present in the
spectrum measured using the signal (\iso{6}{Li} + \iso{3}{He}) reaction
which was not present in the background (\iso{6}{Li} + \iso{4}{He})
measurement. We hypothesize that this additional background resulted from positrons
emitted from \bor  implanted on surfaces other than the
silicon detector.  These positrons may pass through both the silicon and
scintillator detectors and result in a false coincidence trigger and
a continuum of events in the energy spectrum.

Since this background had the same general shape as the 
larger background observed with the beam off or with \iso{4}{He} in the
gas cell, we assume it is composed primarily of a spectrum of the same
shape, but we will allow for the possibility of additional background 
in the energy range of interest.  Such a background would be identified
by additional counts in the spectrum adjacent to the energy window for the
ground state decay.

\begin{figure}[h]
\medskip
\includegraphics[ width=.48 \textwidth]{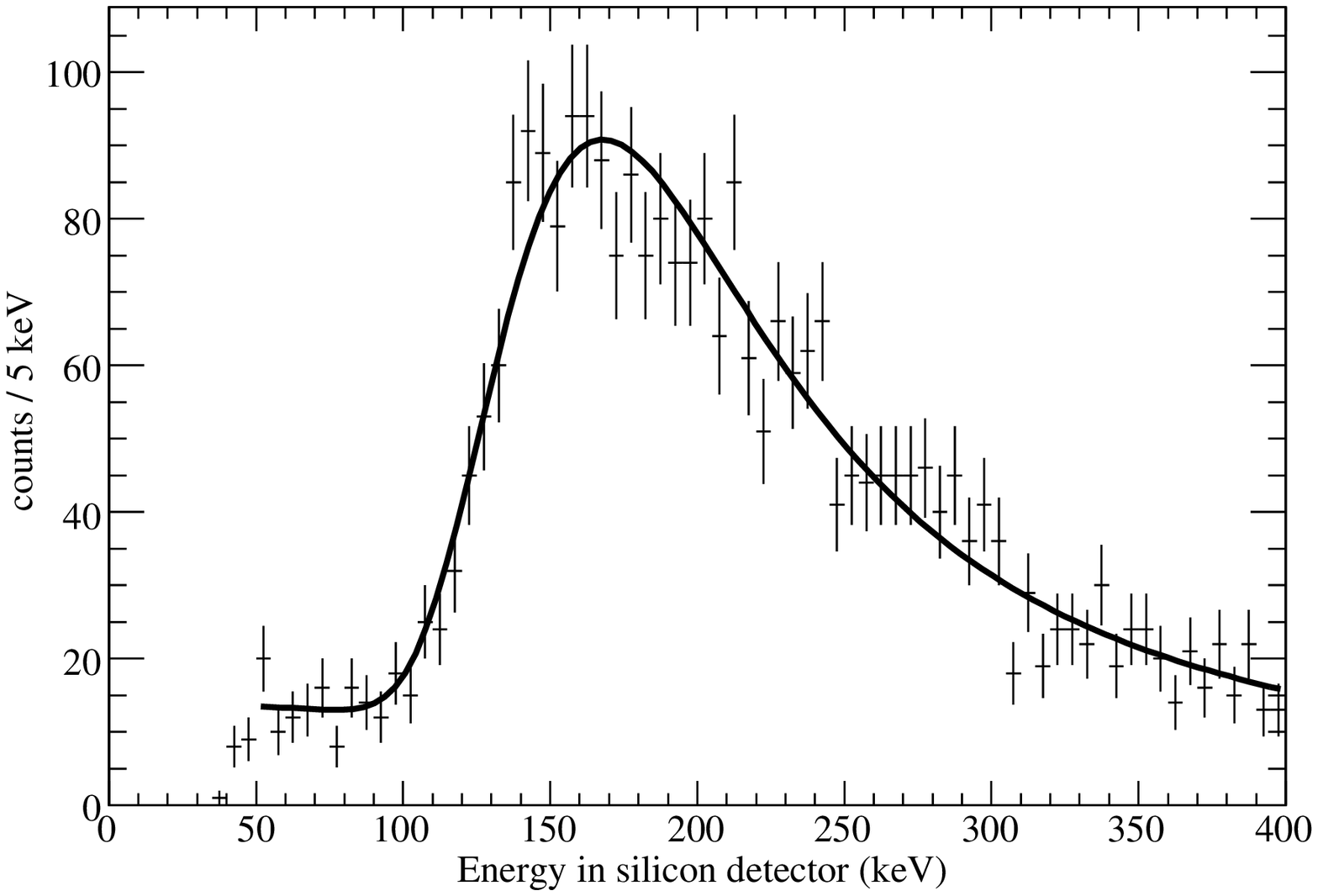}
\bigskip

\includegraphics[  width=.48 \textwidth]{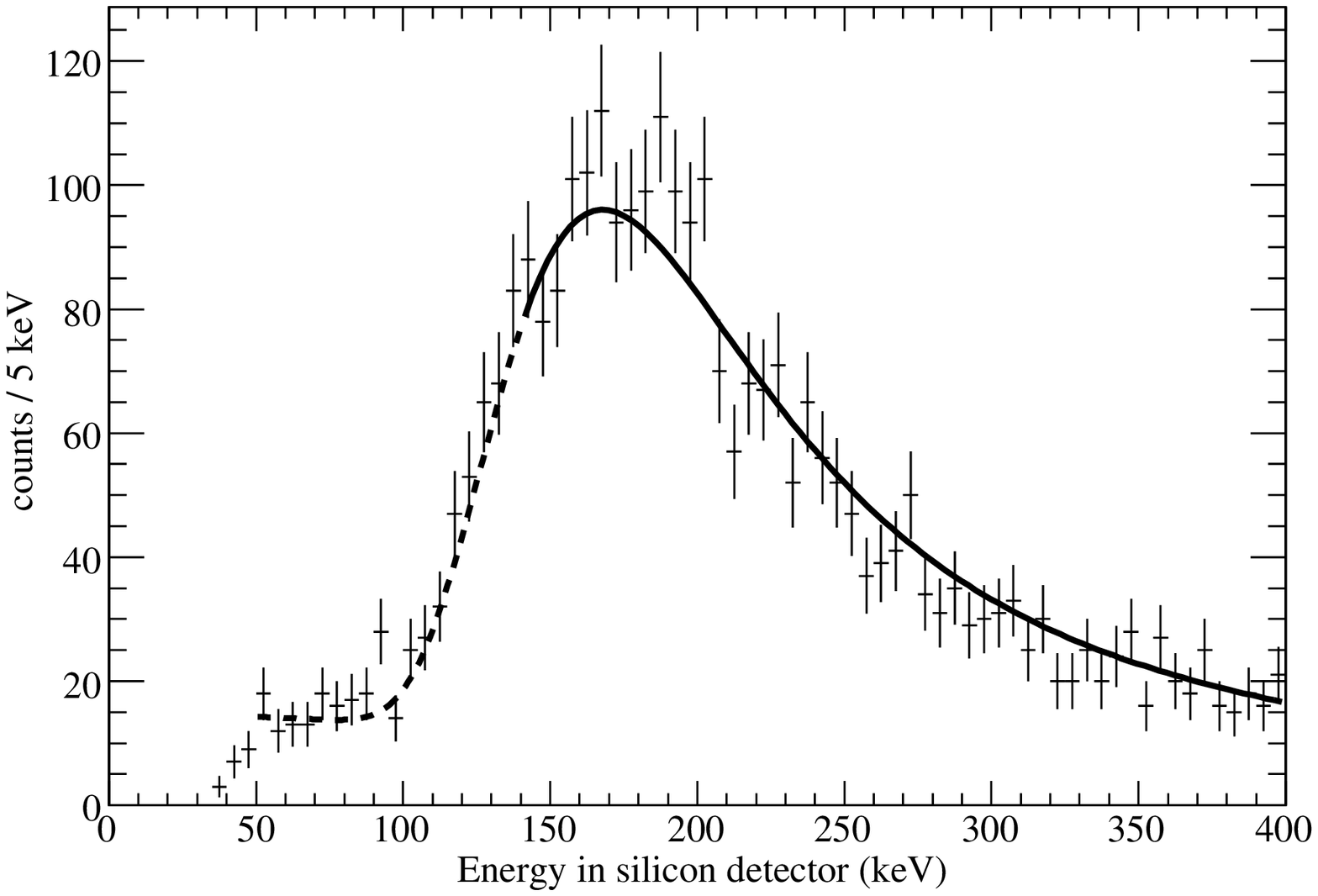}

\caption{\label{fig:run5_sigbkg} Data from the second experimental
  run. The top plot is the data measured with \iso{4}{He} in the gas cell,
  with the fit to this background.    The bottom plot is the 
  data measured with \iso{3}{He} in the gas cell, with a fit to the 
  data above 140~keV 
  using the parameters obtained from the
  background spectrum, but allowing for an adjustable multiplicative
  normalization. The dashed range of the second curve denotes the
  energy region of interest for signal extraction.}

\end{figure}

In order to account for the first part of this background, we parameterized the shape of
the background spectrum by fitting the measured background energy
spectrum from $50-400$~keV, as shown in the first plot of
Fig.~\ref{fig:run5_sigbkg}. The parameterization function consisted of
a Landau distribution and a linear term. 
Holding the parameters
from the background fit fixed and allowing for a free multiplicative
normalization factor, the parameterization function was then fit to the
signal energy spectrum over an energy window from $140-400$~keV, as
shown in the second plot of Fig.~\ref{fig:run5_sigbkg}. 
Excluding the
area around the \gs decay signature window ensures that the normalization
was not influenced by counts in that window. The
number of expected background events in the signal region was obtained
by integrating the fit function over the energy region of
interest. In order to identify the second part of this background,
the excess of counts above this calculated background was
computed in the signal window ($75-120$~keV) and two adjacent energy
windows ($50-75$~keV and $120-140$~keV).

The excess of net counts above the fit background in all three energy windows
confirms the second background, which is not present
in the \iso{4}{He} measurements from which the initial background estimates
are derived. To account for this additional background, we  
interpolated between the two adjacent windows to determine the number
of background events in the primary energy window. We   then
subtracted this background from the total number of excess events
obtained from the difference between the data and the fit, to get the
net excess of signal events. To account for the loss of efficiency due
to positron backscattering, we multiplied the net excess of counts by
1.13. 
The branching ratio was calculated by dividing the corrected net
excess number of counts by the total number of detected coincident
events. The results of this analysis are given in Table
\ref{tab:landexp_sig_fit}.

\begin{table*}[htdp]                                                                                
\begin{center}
\caption{Number of measured events 
 and predicted background in the spectrum obtained with \iso{3}{He} gas in 
 the cell.  Both experimental runs are combined.
}
\label{tab:landexp_sig_fit}
\begin{ruledtabular}
\begin{tabular}{cdddcc}
energy window (keV) &\multicolumn{1}{c}{ events from data\hspace*{3em}} & \multicolumn{1}{c}{ bkg.~events from
fit} & \multicolumn{1}{c}{\hspace*{2em} net remaining} & interpolated bkg.& net excess\\
\hline
50--75 & 282\pm16.8 & 255.7\pm16.0 & 26.2\pm23.2 &---& --- \\ 
75--120 & 661\pm25.7 &575.4\pm24.0 & 85.6\pm35.2&$40.4\pm44.4$ &$45.2\pm56.7$ \\ 
120--140 & 589\pm24.2 & 573.9\pm23.8 & 15.1\pm34.0 &---&---\\ 
\end{tabular}
\end{ruledtabular}
\end{center}
\end{table*}

The data from the two experimental runs, given in table \ref{tab:landexp_sig_fit},
combined with the total of $1.96 \times 10^{6}$ decays collected,
yields a
branching ratio for the \gs decay of $(2.6\pm3.3)\times 10^{-5}$. This
result   is
consistent with no measured signal and corresponds to an
upper limit on the branching ratio of $7.3\times10^{-5}$ at a 90\%
 confidence level.

\section{\label{sec:theory}
Theoretical estimate of branching ratio}

One published estimate \cite{ortizthesis, garciapriv}  of the \gs decay 
rate is based on conserved
vector current (CVC), which relates the vector contribution to the
beta-decay rate to the gamma decay of the 16-MeV isobaric analogue
state of \iso{8}{Be} \cite{DeBraeckeleer:1995aa}. These calculations
do not consider the contribution of the axial matrix element to the
decay rate. Another estimate \cite{tribbleest} 
is based on several different model calculations of the axial-vector 
second-forbidden contribution to the decay to the first excited state of
\iso{8}{Be}.  Those authors suggest that, based on the similarity 
of the intrinsic wave functions of the ground and first excited states
of \iso{8}{Be}, these calculations can be applied to the ground state
decay, and they find a branching ratio of approximately $3 \times 10^{-6}$.
However, since there is no experimental information available to
constrain the contribution of the axial matrix element to the
transition, a strong axial contribution cannot simply be ruled out.

In this section, we discuss our application of the framework developed
by Walecka \cite{Walecka:1975wf, Walecka:2004xd} and Donnelly and
Haxton  \cite{Donnelly:1979uj} for calculating the rates of
electromagnetic and semileptonic processes. The details of the
calculations discussed in this section are specific to the \gs decay.
Our application of this framework is described in detail in Ref.~\cite{mineshthesis}.

\subsection{\label{subsec:shell}Shell model estimate}
The rate of the \gs second-forbidden \bor decay  can be expressed in terms of
kinematic variables and multipole matrix elements, using the formalism in  \cite{Walecka:1975wf}. 
This formalism is quite general,
as it includes all multipoles and leaves the local weak current unspecified --
so that exchange-current or other corrections to the usual one-body operators
can be included. The
transition of interest involves an angular momentum change from J=2 to
J=0, with no change in parity.  The relevant operators, expanded in the long-wavelength
limit (since $q R << \hbar c$), are 
given in terms of the vector and axial vector currents, $\mathbf{J}(\mathbf{x})$
and $\mathbf{J}_5(\mathbf{x})$ 
\cite{Walecka:2004xd}.  These operators, 
$\hat{M}_2(q )$, $\hat{T}_2^\mathrm{el}(q )$, and $\hat{T}_2^\mathrm{mag5}(q )$,
are the rank-two charge and transverse electric projections of the vector
current and the transverse magnetic projection of the axial current, respectively.  

The vector current
appearing in $\hat{T}_2^\mathrm{el}$ has a one-body part and two-body exchange-current corrections, both of which are of the order $(v_\text{nuc}/c)$.  A direct calculation of the important exchange current corrections
would be difficult to do precisely in a model calculation.  Fortunately, under CVC and in the long wavelength limit, it is possible to rewrite matrix elements of $\hat{T}^\mathrm{el}_2$ in terms of those of
$\hat{M}_2$ through Siegert's Theorem \cite{Siegert:1937ov}.
The benefit of this substitution is that the one-body charge operator is of order $(v_\text{nuc}/c)^0$, while
two-body corrections to this operator are of order $(v_\text{nuc}/c)^2$.  Thus Siegert's theorem allows one to
deal with a simpler operator that can be well approximated by its one-body form.  The beta-decay
rate then can be written in terms of  $\hat{M}_2(q )$  and $\hat{T}_2^\mathrm{mag5}(q )$.
As discussed below, the matrix element of the Coulomb operator can be taken from the isobaric
analogue gamma decay transition.  Thus the nuclear-structure dependence of our result is
effectively isolated in a single matrix-element ratio, $\langle \hat{T}^\mathrm{mag5}_2\rangle/\langle \hat{M}_2\rangle$.

\subsection{Calculation of the density matrix}
\label{sec:density_matrix}

 For weak and electromagnetic interactions, the
interaction can be well-approximated by considering only the one-body
interaction \cite{Donnelly:1979uj}.  
We define $\rvert\alpha\rangle$ and $\rvert\beta\rangle$ as complete single-particle
wavefunctions, 
and we also define $\rvert|\alpha|\rangle$ and $\rvert|\beta|\rangle$ as
the reduced (in angular momentum and isospin) forms of these
single-particle wavefunctions, i.e. $\rvert n, l, j, t \rangle$. 
 We can then
write a many-body multipole reduced matrix element $\langle J_fT_f
\lVert \hat{\mathcal{O}}_{JT} \rVert J_i T_i \rangle $ as the sum of
single particle matrix elements $\langle |\alpha| \lVert
\mathcal{O}_{JT} \rVert |\beta| \rangle$ multiplied by 
the density matrix elements $\psi_{\alpha \beta}$, as defined in \cite{Donnelly:1979uj}

For the \gs transition, the coefficients for the density matrix were
obtained from p-shell wavefunctions calculated using the {\scshape glasgow} 
shell-model \cite{Whitehead:1977qp} with the Cohen-Kurath (8-16)2BME
interaction \cite{Cohen:1965qa}. The numerical values are given in
Table \ref{tab:density_matrix_elements}.

\begin{table}[h]
\begin{center}
\caption[Calculated density matrix coefficients for \gs decay]{
  Calculated 
  density matrix elements for
  1p-shell transitions for \gs decay, using the Cohen-Kurath
  (8-16)2BME interaction \cite{Cohen:1965qa} and calculated using code
  obtained from \cite{W.C.-Haxton:iu}. The notation $a
  \rightarrow b$ denotes the density matrix element $\psi_{ab}$, which
  connects the single particle states with $j_i = a$ and $j_f = b$.}
\label{tab:density_matrix_elements}
\begin{ruledtabular}
\begin{tabular}{cd} 
 transition  & \multicolumn{1}{c}{density matrix coefficient}\\
\hline 
    $\frac{1}{2}\rightarrow \frac{3}{2}$   &  -0.206564 
  \\  $\frac{3}{2}\rightarrow \frac{1}{2}$  & +0.168713
  \\ $\frac{3}{2}\rightarrow \frac{3}{2}$& -0.608563
\\
 \end{tabular}
\end{ruledtabular}
\end{center}
\end{table}%

\subsection{Single-particle matrix elements}
\label{sec:spme}
The single-particle operators relevant to this calculation are defined
in terms of spherical Bessel functions $j_J$ and scalar and vector
spherical harmonics $Y_J^{M_J}$ and
$\boldsymbol{\mathcal{Y}}_{JL1}^{M_J}$ \cite{Donnelly:1979uj}:

\begin{align}
M_J^{M_J}(q\mathbf{x})& \equiv j_J(q x)Y_J^{M_J}(\Omega_x),
\\ 
\Sigma_J^{M_J}(q\mathbf{x}) & \equiv  j_J(q
x)\boldsymbol{\mathcal{Y}}_{JJ1}^{M_J}(\Omega_x) \cdot
\boldsymbol{\sigma}.
\end{align}

The matrix elements of these operators can be expanded in a harmonic
oscillator basis and expressed as functions of the form
\cite{Donnelly:1979uj}:
\begin{equation}
\label{eq:DH_spme}
\langle n_f (l_f \,{\scriptstyle \frac{1}{2}})j_f \lVert
T_J(q\mathbf{x}) \rVert n_i
(l_i\,{\scriptstyle\frac{1}{2}})j_i\rangle = \sqrt{\frac{1}{4\pi}}
y^{(J-2)/2}e^{-y}p(y),
\end{equation}
where $T_J(q \mathbf{x})$ is a generic single-particle operator
reduced in spin/isospin, $y \equiv (\frac{1}{2}q b)^2$, b=1.787 fm is
the oscillator parameter \cite{Haxton:PRL41}, and $p(y)$ is a polynomial of 
finite order in y, as tabulated in  \cite{Donnelly:1979uj}.

Folding the matrix elements of \cite{Donnelly:1979uj} with the Cohen-Kurath density matrix yields
\begin{subequations}
\label{eq:dh_eval}
\begin{align}
\label{eq:dh_m_eval}
\langle J_f \lVert \hat{M}_2 \rVert J_i\rangle &= \frac{-0.23 F_1}
{\sqrt{4\pi}} \bigg(-\frac{4y}{3}\bigg) \exp(-y),\\
\label{eq:dh_tm5_eval}
\langle J_f \lVert \hat{T}^\mathrm{mag5}_2 \rVert J_i\rangle
&= \frac{0.04 F_A \sqrt{6}}{\sqrt{4\pi}}\bigg(-\frac{2y}{3}\bigg) \exp(-y),
\end{align}
\end{subequations}
where $F_1=1.0$ and $F_A=1.26$ are the single-nucleon charge and axial-vector couplings.

While the $\hat{M}_2$ and $\hat{T}^{\mathrm{mag5}}_2$ matrix elements
have identical momentum dependence, a suppression of the
$\hat{T}^{\mathrm{mag5}}_2$ matrix element is caused by the
cancelation to 20\% in the sum of the density matrix elements:
\begin{equation}
\psi_{\frac{3}{2}\frac{1}{2}} + \psi_{\frac{1}{2}\frac{3}{2}} = 0.17 -
0.21 = -0.04.
\end{equation}
The uncertainty of this model dependence is difficult to quantify, and
therefore we unfortunately cannot be assured that this result reflects
the true strength of the axial contribution to this decay. Despite
this, we are now able 
to obtain the beta decay
rate for the \gs decay.  We find that the vector contribution dominates,
and the resulting branching ratio is $6.1 \times 10^{-5}$.

\subsection{\label{subsec:vector}Vector contribution derived from
  electromagnetic rate}

Since our shell model calculation showed dominance of the vector contribution to the
decay, it is useful to compare that result with the value implied by measurements
of isobaric analogue electromagnetic transitions.
The doublet in \iso{8}{Be} at 16.6/16.9 MeV is composed of a
$T=0$ state mixed with the $T=1$ isobaric analogue state of the \bor ground
state. The reduced strengths for the E2 transition from these states
to the ground state of \iso{8}{Be} are $B(E2)_{16.6} = 0.068 \pm
0.024\,e^2\,\mathrm{fm^4}$ and $B(E2)_{16.9} = 0.075\pm0.013 \,
e^2\,\mathrm{fm^4}$ \cite{DeBraeckeleer:1995aa}. Due to ambiguity in
the phase of the matrix elements, decomposing the measured transition
strengths into isovector and isoscalar components gives two results:
either $B(E2)_\mathrm{IV} = 0.00\pm0.03 \, e^2\,\mathrm{fm^4}$ and
$B(E2)_\mathrm{IS} = 0.14 \pm 0.03 \,e^2\,\mathrm{fm^4}$, or
alternatively, $B(E2)_\mathrm{IV} = 0.14\pm0.03 \,e^2\,\mathrm{fm^4} $
and $B(E2)_\mathrm{IS} = 0.00 \pm 0.03 \,e^2\,\mathrm{fm^4}$. We will
henceforth use the maximum ($0.14+0.03=0.17 \,e^2\,\mathrm{fm^4}$) and
minimum (0.0) values for $B(E2)_\mathrm{IV}$ to constrain the beta
decay transition rate.

The rate of the nuclear electromagnetic transition rate in terms of
the multipole operators defined in the previous section is \cite{Walecka:2004xd},
\begin{equation}
\label{eq:full_gamma_rate}
\omega_{\gamma} = \frac{8\pi\alpha k_\gamma}{2J_i + 1}  |\langle
  f \lVert \hat{T}^\mathrm{el}_J \rVert i \rangle|^2, 
\end{equation}  
where $\omega_{\gamma}$ is the transition rate per unit time,
$k_\gamma$ is the gamma-ray momentum, $\alpha$ is the fine structure
constant, $J$ is the angular momentum change in the transition, and
$J_i$ is the initial spin of the nucleus. As with the beta decay
expression, we can rewrite this in terms of the $\hat{M}_2$ matrix
element using Siegert's Theorem \cite{Siegert:1937ov} and substituting $J=J_i=2$,
\begin{equation}
\label{eq:gamma_rate}
\omega_{\gamma} =
\frac{3}{2}\frac{8\pi\alpha}{5}\frac{(E_i-E_f)^2}{k_\gamma} |\langle f
\lVert \hat{M}_2 \rVert i \rangle|^2.
\end{equation}
The energy released in the isovector decay is 16.80 MeV
\cite{DeBraeckeleer:1995aa}, and the corresponding decay rate from
Eq.~\ref{eq:gamma_rate} yields a $B(E2)$ value of 0.55 $e^2$ fm$^4$
for the transition, using the ${\hat{M}}_2$ matrix element obtained in
the previous section. In order for this calculation to match the
measured $B(E2) = 0.17$ $e^2$ fm$^4$, the matrix element must be
reduced by a factor of 1.8.  To incorporate the second
($B(E2)_\mathrm{IV} = 0$) constraint, we simply set $M_2 = 0$ in the
beta decay calculation.

If we assume the maximum possible strength of the vector matrix
element from the gamma decay measurement, we obtain a transition rate
of $1.8\times 10^{-5} \ \text{s}^{-1}$. On the other hand, if we assume that
the vector matrix element does not contribute, the transition rate is
reduced to $9.1\times 10^{-7}\ \text{s}^{-1}$. The respective branching
ratios for these cases are $2.0\times10^{-5}$ and $1.0\times10^{-6}$.

\section{\label{sec:conclusion}Conclusions}

An accurate measurement of or limit on the \gs beta decay branching ratio is of 
importance for the current generation of solar neutrino
measurements.  
To the best of our knowledge, this work is the first
published attempt at the measurement of the \gs decay branching ratio.

Our measurement of the \gs beta decay transition centers on the
detection of the two low-energy alpha particles from the breakup of
\iso{8}{Be}. A significant challenge to this measurement is the
characterization of the response of our silicon detector to low-energy
alpha particles. In light of this, we have done two experiments to
determine our detector response.  The first measurement involved
directly implanting alpha particles with kinetic energy between 
86 and 453 keV
into our detector, while simultaneously measuring the detector
deadlayer. The second measurement involved the implantation of
\iso{12}{N}, which is unstable to beta-delayed $3\alpha$ decay. The results from
both of these measurements verify that the tabulated nuclear stopping
powers for alpha particles in silicon are reasonably valid for the
range of energies we are considering. However, the directly-implanted
low-energy alpha particle measurements indicates that the uncertainty
on the tabulated values may be larger than predicted.

Using the measurements and techniques described above, we have
measured the \gs decay branching ratio to be $(2.6\pm3.3)\times 
10^{-5}$. This measurement is consistent with zero and can be
used to place an upper limit on the branching ratio of
$7.3\times10^{-5}$ at the 90\% confidence level. 
At this level, the \gs decay branch is not a
significant background to measurements of the spectral shape of solar
\iso{8}{B} neutrinos and the solar hep neutrino flux. Our
calculation of the \gs decay rate gives a  
branching ratio of $6.1\times 10^{-5},$ using values for the vector and 
axial-vector matrix elements obtained from the shell model.  
The model-dependent axial term in our calculation
may be quite uncertain. We find a branching ratio range of
$1.0\times10^{-6}$ to $2.0\times 10^{-5}$, based on the values for the vector matrix element
obtained from the experiment of Ref.~\cite{DeBraeckeleer:1995aa}  
along with the axial-vector matrix element obtained from the shell model.
The values are all consistent with our measured value.

\begin{acknowledgments}

The authors would like to thank Wick Haxton for his assistance with
the theoretical aspects of this work. The CENPA technical staff: John
Amsbaugh, Greg Harper, Mark Howe, and Doug Will, also contributed
extraordinary amounts of time and effort in support of the
experimental measurements. We also wish to thank Eric Adelberger,
Manojeet Bhattacharya, Jason Detwiler, Alejandro Garcia, Seth Hoedl,
and John Wilkerson for their wealth of ideas and insights. Finally,
work by Matthias Gohl, Shuje Uehara, Ashley Batchelor, Christy McKinley, Patrick Peplowski 
and many other UW Tandem crew members, was crucial to 
carrying out this work. This research was partially supported by DOE Grant DE-FG03-97ER41020.
\end{acknowledgments}
\bibliography{b8_gs}
\end{document}